\documentclass[apl,twocolumn,superscriptaddress]{revtex4}
\usepackage{bm,graphicx,graphics,amsmath,amssymb,bm,epsfig,color}
\usepackage{euscript,tabularx}
\usepackage{textcomp}
\usepackage{gensymb}
\usepackage{longtable}

\begin{document}
\bibliographystyle{apsrev}

\title{A Frequency-Controlled Magnetic Vortex Memory}

\author{B. Pigeau}
\affiliation{Service de Physique de l'\'Etat Condens\'e (CNRS URA 2464), CEA Saclay, 91191 Gif-sur-Yvette, France}

\author{G. de Loubens}
\thanks{Corresponding author: gregoire.deloubens@cea.fr}
\affiliation{Service de Physique de l'\'Etat Condens\'e (CNRS URA 2464), CEA Saclay, 91191 Gif-sur-Yvette, France}

\author{O. Klein}
\affiliation{Service de Physique de l'\'Etat Condens\'e (CNRS URA 2464), CEA Saclay, 91191 Gif-sur-Yvette, France}

\author{A. Riegler}
\affiliation{Physikalisches Institut (EP3), Universit\"at W\"urzburg, 97074
  W\"urzburg, Germany}

\author{F. Lochner}
\affiliation{Physikalisches Institut (EP3), Universit\"at W\"urzburg, 97074
  W\"urzburg, Germany}

\author{G. Schmidt}
\thanks{Present address: Institut f\"ur Physik,
  Martin-Luther-Universit\"at, Halle Wittenberg, 06099 Halle, Germany}
\affiliation{Physikalisches Institut (EP3), Universit\"at W\"urzburg, 97074
  W\"urzburg, Germany}

\author{L.~W. Molenkamp}
\affiliation{Physikalisches Institut (EP3), Universit\"at W\"urzburg, 97074
  W\"urzburg, Germany}

\author{V.~S. Tiberkevich}
\affiliation{Department of Physics, Oakland University, Michigan 48309, USA}

\author{A.~N. Slavin}
\affiliation{Department of Physics, Oakland University, Michigan 48309, USA}

\date{\today}

\begin{abstract}
  Using the ultra low damping NiMnSb half-Heusler alloy patterned into
  vortex-state magnetic nano-dots, we demonstrate a new concept of
  non-volatile memory controlled by the frequency. A perpendicular
  bias magnetic field is used to split the frequency of the vortex
  core gyrotropic rotation into two distinct frequencies, depending on
  the sign of the vortex core polarity $p=\pm1$ inside the dot. A
  magnetic resonance force microscope and microwave pulses applied at
  one of these two resonant frequencies allow for local and
  deterministic addressing of binary information (core polarity).
\end{abstract}

\maketitle

One of the most important goals of the modern information technology
is the development of fast high-density non-volatile random access
memories (RAM) that are energy efficient and can be produced using
modern planar micro- and nano-fabrication methods. Magnetic
nano-objects offer a convenient way to store binary information
through their bi-stable properties, but the development of practical
magnetic RAM requires to find a performant mechanism to reverse the
magnetization inside individual cells \cite{schumacher03}. One of the
ways is to take advantage of the high dynamical susceptibility of
magnetic nano-objects at their ferromagnetic resonance frequency. This
resonance approach is particularly efficient for low dissipation
materials, as it allows to concentrate the energy in a narrow
frequency band and to reduce the energy cost of the reversal process
by the quality factor of the resonance.

In a vortex-state magnetic nano-dot \cite{guslienko08}, the static
magnetization is curling in the dot plane, except in the dot center
where it is forming an out-of-plane vortex core \cite{shinjo00} of
typical size of the exchange length $l_\text{ex}\simeq$ 5-10~nm. The
core can be directed either perpendicularly up or down relative to the
dot plane, this bi-stability being characterized by the core polarity
$p=\pm 1$. Recent experiments demonstrated that the vortex core
polarity can be reversed in \emph{zero} applied magnetic field through
the excitation of the gyrotropic rotation of the vortex core about its
equilibrium position at the dot
center \cite{waeyenberge06,yamada07}. At $H=0$, the frequency of this
gyrotropic core rotation mode, $f_0$, is identical for both core
polarities, but the sense of this rotation depends on $p$. Thus, for a
given core polarity, the \emph{circular polarization} of the microwave
field -- right or left depending on the sign of $p$ -- discriminates
the occurrence of the resonant microwave absorption by the
vortex-state magnetic dot \cite{curcic08}.  When the radius $r$ of the
core orbit increases, a distortion of the core profile characterized
by the appearance of a tail having the magnetization direction
opposite to that of the original core polarity
occurs \cite{yamada07,guslienko08a,vansteenkiste09}. The magnitude of
this tail depends solely on the linear velocity $V=2\pi f_0 r$ of the
vortex core \cite{guslienko08a}. When the latter reaches the critical
value $V_c = (1/3) \omega_M l_\text{ex}$ at $H=0$ (where
$l_\text{ex}=(2A_\text{ex}/\mu_0 M_0^2)^{1/2}$, $\omega_M=\gamma \mu_0
M_0$, $\mu_0$ is the permeability of the vacuum, $M_0$ the saturation
magnetization of the magnetic material, $\gamma$ its gyromagnetic
ratio, and $A_\text{ex}$ its exchange constant), the core polarity
suddenly -- within few tens of picoseconds \cite{lee08a,hertel06} --
reverses.

Still, \emph{reliable} control of an \emph{individual} cell in a large
array based on resonant switching, which takes full advantage of the
frequency selectivity of magnetic resonance, has to be realized. In
this letter, we show that a frequency-controlled memory with resonance
reading and writing schemes can be realized using vortex-state NiMnSb
nano-dots placed in a perpendicular magnetic bias field $H \neq 0$. In
our experimental realization, local adressing of the core polarity is
achieved by means of a magnetic resonance force microscope (Fig.1).
We also propose a more practical variant based on a full solid-state
design.

The key role of the static magnetic field $H$ aligned along the axis
of the vortex core -- $H$ being the sum of an homogeneous bias and of
a small additional local component -- is to introduce a controlled
splitting of the frequency of the gyrotropic mode depending on the
core polarity \cite{loubens09}. Therefore, the polarity state of
individual magnetic dots can be selectively addressed by controlling
the \emph{frequency} of a \emph{linearly polarized} microwave pulse
excitation. When the core polarity is equal to $p = +1$ (i.e.  the
core is parallel to $\textbf{H}$), the rotational frequency $f_+$ is
larger than the frequency $f_-$ corresponding to the core polarity $p
=- 1$ (i.e. the core is antiparallel to $\textbf{H}$). This frequency
splitting is directly proportional to $H$ \cite{ivanov02,loubens09}
(see Fig.2a):
\begin{equation}
  f_+(H)-f_-(H) = 2 f_0 \frac{H}{H_s},
\label{splitting}
\end{equation}
where $H_s$ is the magnetic field required to saturate the dot along
its normal and $f_0$ is the frequency of the gyrotropic mode at $H=0$,
which can be approximated by the analytical
expression \cite{guslienko08}: $f_0 = (10/9) (\omega_M/2\pi) \beta$,
where $\beta=t/R$, $t$ is the thickness and $R$ the radius of the dot.

\begin{figure}
  \includegraphics[width=5cm]{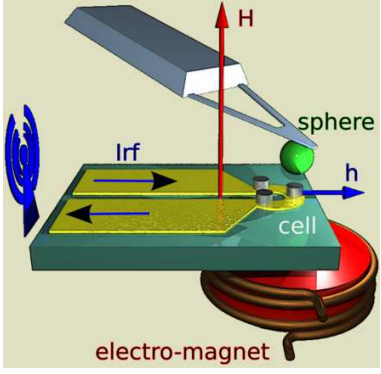}
  \caption{Prototype of a frequency-controlled magnetic memory
    realized by means of a magnetic resonant force microscope. The
    memory elements are vortex-state NiMnSb disks of diameter $1~\mu$m
    and thickness $44$~nm separated by 10~$\mu$m.}
\end{figure}

To design a practical memory cell it is necessary to choose the static
magnetic field $H$ in such a way that the field-induced gyrotropic
frequency splitting Eq.(\ref{splitting}) exceeds the linewidth $\Delta
f$ of the gyrotropic mode which can be approximately expressed as
$\Delta f \simeq \alpha_v f_0$, where $\alpha_v
=\alpha[1+\ln{(R/R_c)}/2]$ is the damping parameter for the gyrotropic
core rotation mode \cite{thiele73,guslienko06}, $\alpha$ is the
dimensionless Gilbert damping constant of the dot magnetic material
and $R_c\sim l_{ex}$ is the vortex core radius. Thus, the minimum
perpendicular bias field is given by the expression:
\begin{equation}
  H_\text{min} \simeq \frac{\alpha_v}{2} H_s
\label{hmin}
\end{equation}
(see Fig.2a). It follows from Eq.(\ref{hmin}), that to reduce
$H_\text{min}$, it is necessary to choose the dot magnetic material
with low damping and to increase the aspect ratio $\beta$ of the dot,
as this leads to the decrease in the saturation field $H_s$. To get a
sufficient frequency separation between the modes with opposite core
polarities it is convenient to have the magnitude of $H$ that is
several times larger than $H_\text{min}$.

\begin{figure}
  \includegraphics[width=7cm]{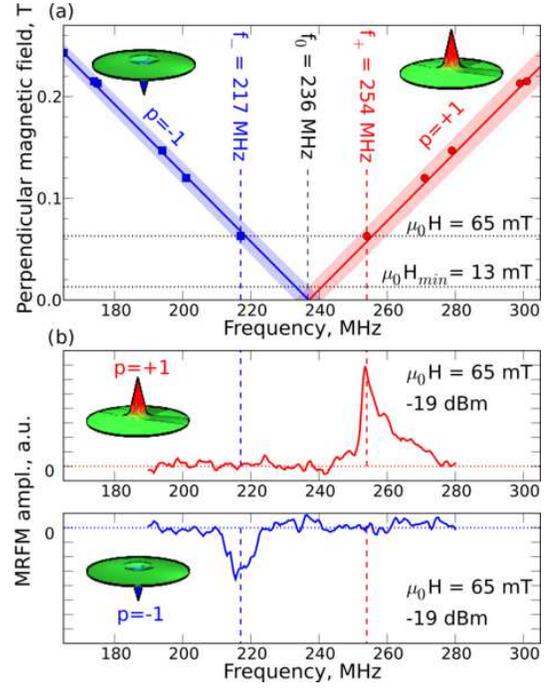}
  \caption{ (a) Frequency splitting induced by a perpendicular
    magnetic field between the gyrotropic modes corresponding to the
    two opposite core polarities $p=\pm1$. The shaded area illustrates
    the broadening associated with the linewidth of the gyrotropic
    mode. (b) The red and blue traces are the MRFM absorption signals
    measured at $\mu_0 H=65$~mT for $p=\pm1$.}
\end{figure}

The design of our experimental frequency-controlled magnetic vortex
memory is presented in Fig.1. The memory elements are circular
magnetic dots made of an epitaxial, ultra low damping half-Heusler
alloy with high Curie temperature, NiMnSb(001) ($\alpha=0.002$, $\mu_0
M_0=690$~mT, $T_C=730$~K) \cite{bach03, heinrich04}. Their aspect ratio
$\beta\simeq 0.1$ ($t=44$~nm, $R=500$~nm) is relatively large, and
they are separated from each other by 10~$\mu$m.  The detailed
magnetic characterization of such NiMnSb dots was performed in
Ref. \cite{loubens09} and yields the saturation field $ \mu_0
H_s=800$~mT. An electromagnet is used to produce a tunable
perpendicular magnetic field homogeneous on all the dots, and oriented
perpendicular to the plane. This static field creates the above
mentioned splitting of the gyrotropic frequencies for different core
polarities. The dots are placed at the extremity of an
impedance-matched gold microwave strip-line which provides an in-plane
linearly polarized microwave magnetic field $\textbf{h}$. Since
$\textbf{h}$ contains both right and left circularly polarized
components, it couples to the gyrotropic rotation of the vortex core
for both core polarizations $p=\pm1$. This microwave field with
variable frequency $f$ is used to resonantly excite gyrotropic
rotation of the vortex core in a magnetic dot.

If the microwave field is weak, the amplitude of this gyrotropic
rotation is relatively small, but sufficient to \emph{read} the
polarity of the rotating core (without destroying it) using the
technique of magnetic resonance force microscope (MRFM), which is
illustrated schematically in Fig.1 and described in detail in
Ref. \cite{klein08}. If the microwave field is sufficiently large and
has the frequency corresponding to the resonance gyrotropic frequency
for a given core polarity (e.g. $f_+$ for $p=+1$), the velocity of the
vortex core rotation induced by this field reaches the critical value,
and the core polarity is reversed (\emph{written}).

Achieving a dense memory requires to address the vortex core polarity
state of a selected magnetic nano-dot inside an array. In our
experimental memory prototype of Fig.1, we meet this challenge using
the MRFM technique \cite{loubens09,klein08}. In the framework of this
technique the magnetic probe glued to a soft cantilever is scanned
horizontally over the different magnetic dots. This probe is a 800~nm
diameter sphere made of amorphous Fe (with 3\% Si), and its role is
two-fold. First of all, it works as a sensitive local probe capable of
detecting the change of the vertical ($z$) component of magnetization
of a single magnetic dot caused by the gyrotropic rotation of the
vortex core in this dot. Thus, it is possible to read the polarity of
the vortex core in a selected dot \cite{loubens09} (Fig.2b). Secondly,
the dipolar stray field of the magnetic probe creates an additional
local bias field of about 20~mT (i.e. roughly twice as large as $\mu_0
H_\text{min} \simeq 13$~mT), which allows one to single out the
particular magnetic dot situated immediately under the probe in the
information writing process. The presence of the local bias field
created by the probe shifts the gyrotropic frequency in this dot by
about $\Delta f$, thus allowing one to choose the frequency of the
microwave writing signal in such a way, that the reversal of core
polarity is done in only the selected dot, without affecting the
information stored in the neighboring dots.

\begin{figure}
  \includegraphics[width=8.5cm]{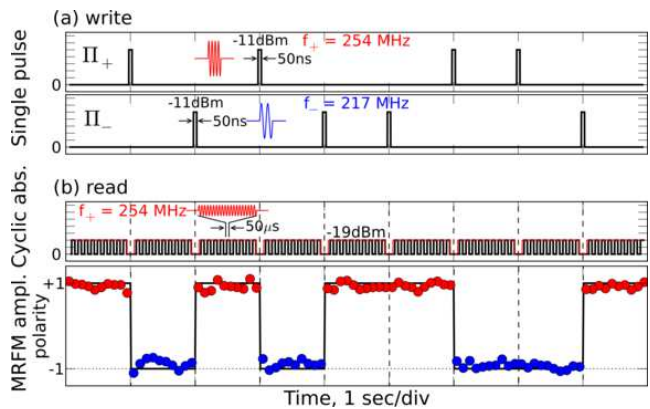}
  \caption{Local frequency control of the binary information
    demonstrated at $\mu_0 H=65$~mT. (a) The writing is performed
    every second by applying a single microwave pulse ($\tau_w=50$~ns,
    $P_w=-11$~dBm) whose carrier frequency is tuned at either $f_+$ or
    $f_-$.  (b) The reading ($P_r=-19$~dBm) is performed continuously
    between the writing pulses by MRFM using a cyclic absorption
    sequence at the cantilever frequency ($\sim$ 10~kHz or half-period
    $50~\mu$s).}
\end{figure}

In our experiments the total static magnetic field was chosen to be
$\mu_0 H=65$~mT $\simeq 5~\mu_0 H_\text{min}$, which gave the
following gyrotropic frequencies of the experimental magnetic dot:
$f_+ =254$~MHz, $f_-=217$~MHz, and $f_0=236$~MHz (Fig.2). The
frequency linewidth of the gyrotropic rotation in the experimental dot
was of the order of $\Delta f \simeq 8$~MHz.

The process of reading of the binary information stored in a magnetic
dot is illustrated by Fig.2b. The amplitude of the cantilever
oscillations is measured while the frequency of the weak (reading)
microwave signal is varied in the interval containing $f_+$ and
$f_-$. The results of these measurements are shown for the cases when
the vortex core polarity was set at $p=+1$ or $p=-1$ at the beginning
of the microwave frequency sweep \footnote{By saturating all the dots
  in a large positive -- or, respectively, negative -- static magnetic
  field.}. It is clear from Fig.2b that the core polarity can be
detected not only from the resonance signal frequency, which is
different for different core polarities, but also from the sign of the
MRFM signal \cite{loubens09}, which is positive for $p=+1$ and
negative for $p=-1$.

The writing process in a dot with initial core polarity equal to
$p=+1$ is illustrated by Fig.3. Fig.3a shows the frequency $f_w$ of
the strong writing pulses of width $\tau_w=50$~ns and power $P_w\simeq
100~\mu$W (corresponding to a microwave magnetic field of $\mu_0
h=0.3$~mT), while Fig.3b shows the frequency $f_r$ of the weak
reading signal of power $P_r \simeq 10~\mu$W, which is supplied
continuously and is interrupted every second in order to apply a
strong writing pulse. The frequency of the weak reading signal can be
kept close to either $f_+$ or $f_-$, and the amplitude of cantilever
oscillations measured by MRFM provides the reading of the core
polarity, as presented in Fig.3b for the $f_+$ case.

The application of the first writing pulse to a particular selected
dot having the initial polarity $p=+1$ results in the excitation of
the vortex core rotation of the resonance frequency $f_{+}=254$~MHz
and the amplitude which is sufficient to bring the vortex core to the
threshold speed corresponding to the core polarity
reversal \cite{guslienko08a}.  Once inverted, the final state $p=-1$ is
out of resonance with the writing pulse (as $f_- < f_+-\Delta f$), so
that the polarity can not be switched back to $p=+1$. It is clear from
Fig.3 that the writing pulses of the carrier frequency $f_+$ (that we
shall call $\Pi_+$-pulses) change the vortex core polarity from $p=+1$
to $p=-1$, while the writing pulses of the carrier frequency $f_-$
($\Pi_-$-pulses) change the core polarity from $p=-1$ to $p=+1$. For
the chosen parameters of the writing pulses the polarity reversal is
deterministic: the reversal efficiency has been tested several hundred
times without any failure, implying a success rate better than
99\%. We also note that the application of the $\Pi_+$-pulse to the
magnetic dot with the polarity $p=-1$ (and application of a
$\Pi_-$-pulse to the dot with $p=+1$) does not have any effect on the
vortex core polarity in the dot. Moving the MRFM probe to the
neighboring dots during the reading sequence allows one to check that
the core polarity in adjacent dots (situated 10~$\mu$m away) is
unaffected by the core reversal process in the selected dot. Thus, it
has been demonstrated that the frequency-selective deterministic
manipulation of the binary information has been achieved locally.

\begin{figure}
  \includegraphics[width=7cm]{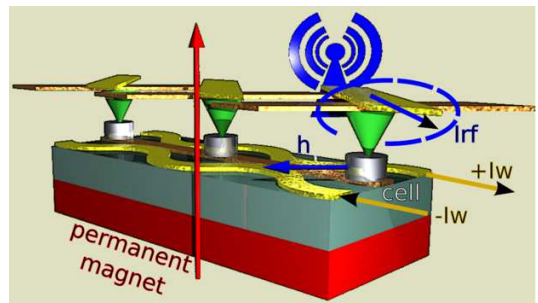}
  \caption{Proposed solid state design of the frequency-controlled
    magnetic memory.}
\end{figure}

Although the experimental device shown in Fig.1 can be used as a
prototype for the development of a frequency-controlled magnetic
memory, a series of improvements can be imagined to make a more
practical solid-state variant (Fig.4). Firstly, it would be useful to
increase the dot aspect ratio to $\beta=t/R=1$ in order to reduce the
dot saturation field $H_s$, and, therefore, the minimum perpendicular
bias magnetic field to $\mu_0 H_\text{min}\simeq 5$~mT (see
Eq.(\ref{hmin})). In this case, a static bias field of only 20~mT,
that could be produced by a permanent magnet placed underneath the
substrate, should be sufficient to ensure reliable operation of the
memory. Secondly, the dots of the practical variant should be arranged
in a regular square array, where addressing of a particular dot is
achieved by local combination of the static and microwave fields at
the intersection of a word and a bit lines. The word line could be
made in the form of a pair of wires running parallel to each row of
dots at a 100~nm separation distance. A bias current $I_w=5$~mA would
be sufficient to create an additional perpendicular field of 10~mT at
the addressed row, causing an additional shift of the resonance
frequency by about a full linewidth. The bit line could be made as an
impedance matched wire running above each column of dots, producing
the in-plane linearly polarized microwave field $h$. Thirdly, it would
be useful to replace the MRFM detection of Fig.1, which contains
mechanically moving parts, by local electrical detectors of the
absorbed power for the information reading process. Finally, we would
like to point out that the proposed design offers the possibility to
create a multi-register memory by stacking dots of different aspect
ratios $\beta$ on top of each other, as they will have different
resonance frequencies of the vortex core rotation.

This research was partially supported by the French Grant Voice
ANR-09-NANO-006-01, by the European Grants DynaMax FP6-IST
033749 and Master NMP-FP7 212257, by the MURI Grant
W911NF-04-1-0247 from the U.S. Army Research Office, by the
Contract W56HZV-09-P-L564 from the U.S. Army TARDEC, RDECOM, and by
the Grant No. ECCS-0653901 from the U.S. National Science Foundation.

\end{document}